\documentclass[manuscript,screen]{acmart}
\usepackage{geometry}
\usepackage{float}
\usepackage[table]{xcolor}
\usepackage{amsmath}

\usepackage{amssymb}
 
\AtBeginDocument{%
  }
 
\setcopyright{acmlicensed}
\copyrightyear{2026}
\acmYear{2026}
\acmDOI{XXXXXXX.XXXXXXX}
\acmISBN{978-1-4503-XXXX-X/2024/XX}
 
\begin{document}
 
\hypersetup{citecolor=black, linkcolor=black, urlcolor=black}
 
\title{TAMA: A Human-AI Collaborative Thematic Analysis Framework Using Multi-Agent LLMs for Clinical Interviews}
 
\author{Huimin Xu}
\authornote{Both authors contributed equally to this research.}
\affiliation{%
  \institution{School of Information, University of Texas at Austin}
  \city{Austin}
  \state{TX}
  \country{USA}}
 
\author{Seungjun Yi}
\authornotemark[1]
\affiliation{%
  \institution{Department of Biomedical Engineering, University of Texas at Austin}
  \city{Austin}
  \state{TX}
  \country{USA}}
 
\author{Terence Lim}
\affiliation{%
  \institution{College of Natural Sciences, University of Texas at Austin}
  \city{Austin}
  \state{TX}
  \country{USA}}
\affiliation{%
  \institution{Graphen, Inc.}
  \city{New York}
  \state{NY}
  \country{USA}}
 
\author{Jiawei Xu}
\affiliation{%
  \institution{School of Information, University of Texas at Austin}
  \city{Austin}
  \state{TX}
  \country{USA}}
 
\author{Andrew Well}
\affiliation{%
  \institution{Department of Cardiac Surgery, Division of Pediatric Cardiac Surgery, Vanderbilt University School of Medicine}
  \city{Nashville}
  \state{TN}
  \country{USA}}
\affiliation{%
  \institution{Pediatric Heart Institute, Monroe Carell Jr. Children's Hospital at Vanderbilt}
  \city{Nashville}
  \state{TN}
  \country{USA}}
 
\author{Carlos Mery}
\affiliation{%
  \institution{Department of Cardiac Surgery, Division of Pediatric Cardiac Surgery, Vanderbilt University School of Medicine}
  \city{Nashville}
  \state{TN}
  \country{USA}}
\affiliation{%
  \institution{Pediatric Heart Institute, Monroe Carell Jr. Children's Hospital at Vanderbilt}
  \city{Nashville}
  \state{TN}
  \country{USA}}
 
\author{Aidong Zhang}
\affiliation{%
  \institution{Department of Computer Science, University of Virginia}
  \city{Charlottesville}
  \state{VA}
  \country{USA}}
 
\author{Yuji Zhang}
\affiliation{%
  \institution{Department of Computer Science, University of Illinois at Urbana-Champaign}
  \state{IL}
  \country{USA}}
 
\author{Heng Ji}
\affiliation{%
  \institution{Department of Computer Science, University of Illinois at Urbana-Champaign}
  \state{IL}
  \country{USA}}
 
\author{Keshav Pingali}
\affiliation{%
  \institution{Department of Computer Science, University of Texas at Austin}
  \city{Austin}
  \state{TX}
  \country{USA}}
 
\author{Yan Leng}
\affiliation{%
  \institution{McCombs School of Business, University of Texas at Austin}
  \city{Austin}
  \state{TX}
  \country{USA}}
 
\author{Ying Ding}
\authornote{Corresponding author.}
\affiliation{%
  \institution{School of Information, University of Texas at Austin}
  \city{Austin}
  \state{TX}
  \country{USA}}
\affiliation{%
  \institution{Dell Medical School, University of Texas at Austin}
  \city{Austin}
  \state{TX}
  \country{USA}}
\email{ying.ding@ischool.utexas.edu}

\renewcommand{\shortauthors}{Xu et al.}

\begin{abstract}
Thematic analysis (TA) is a widely used qualitative approach for uncovering latent meanings in unstructured text data. TA provides valuable insights in healthcare but is resource-intensive. Large Language Models (LLMs) have been introduced to perform TA, yet their applications in high-stakes healthcare settings, particularly for qualitative clinical interview analysis, remain limited. Here, we propose \textbf{TAMA}: A Human-AI Collaborative \textbf{T}hematic \textbf{A}nalysis framework using \textbf{M}ulti-\textbf{A}gent LLMs for clinical interviews. We leverage the scalability and coherence of multi-agent systems through structured conversations between agents and coordinate the expertise of cardiac experts in TA. Using interview transcripts from parents of children with Anomalous Aortic Origin of a Coronary Artery (AAOCA), a rare congenital heart disease, we demonstrate that TAMA outperforms single-agent LLM TA approaches, achieving higher thematic hit rate, coverage, and distinctiveness. TAMA demonstrates strong potential for automated TA in clinical settings by leveraging multi-agent LLM systems with human-in-the-loop integration by enhancing quality while significantly reducing manual workload. The full implementation is publicly available at \url{https://github.com/Charlie-Yi-SJ/TAMA}.
\end{abstract}



\authorsaddresses{Authors' Contact Information: Ying Ding, School of Information, University of Texas at Austin, Austin, TX, USA and Dell Medical School, University of Texas at Austin, Austin, TX, USA; ying.ding@ischool.utexas.edu.}

\maketitle

\section{Introduction}
 
Thematic Analysis (TA), a qualitative approach used to systematically identify, analyze, and interpret patterns\footnote[3]{A pattern refers to a repeated idea, experience, or issue that appears across multiple data points, while a theme is a broader, interpretive insight that captures the significance of those patterns.} within data, 
is widely applied across various fields. 
Employing inductive TA in healthcare can significantly improve patient care~\cite{Osborn2020, Jowsey2021} , medical practices~\cite{Jowsey2021, Copel2023, Braun2024} ,
and policy decisions~\cite{AL-Ruzzieh2024} . 
Traditionally, TA involves the following processes: (1) Researchers immerse themselves in qualitative data (such as interview transcripts) by reading the material multiple times to become thoroughly familiar with its content, (2) generate initial codes (codebook) to identify significant features of the data, and (3) organize into broader themes that capture meaningful patterns within the dataset. To reduce the manual workload involved in this process, single-agent large language models (LLMs) have been shown to have potential for partially automating TA. However, this approach also encounters challenges in scalability~\cite{castellanos2025largelanguage},  consistency~\cite{settanni2025accuracy}, and coherence~\cite{hairston2025automating}. Moreover, TA in healthcare is inherently a resource-intensive endeavor, which inevitably requires expert analysis and ethical approvals~\cite{Wade2007} , which add complexity and prolong data collection. 
 
\begin{figure}[H]
  \centering
    \includegraphics[width=1\textwidth]{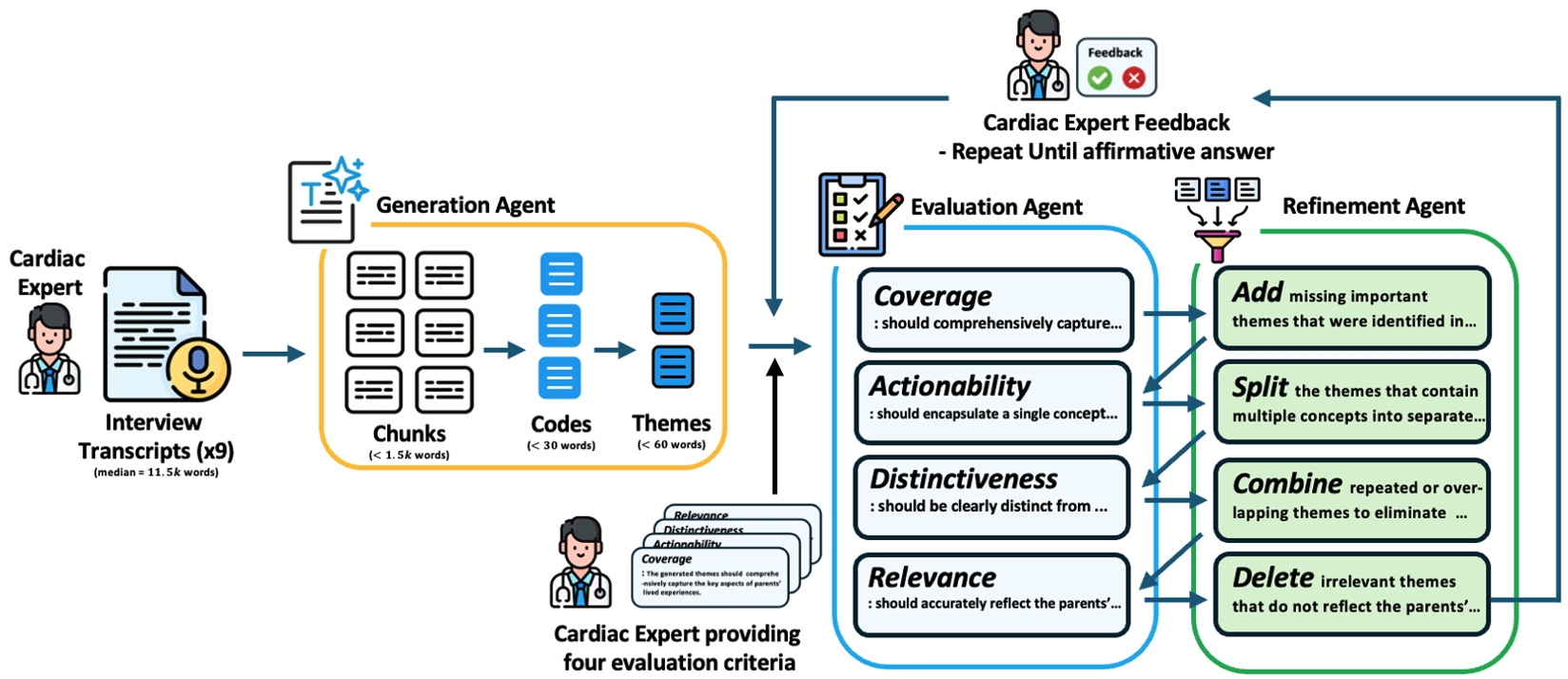}
 \caption{\textbf{The TAMA Framework} is a human-in-the-loop, multi-agent system designed to generate, evaluate, and refine themes from clinical interview transcripts.}
 \label{fig:workflow}
\end{figure}

To address these limitations, we present \textbf{TAMA}: A Human-AI Collaborative \textbf{T}hematic \textbf{A}nalysis framework using \textbf{M}ulti-\textbf{A}gent LLMs for clinical interviews (Figure~\ref{fig:workflow}) to enhance the scalability, consistency, and coherence of inductive TA in healthcare contexts. Using TAMA, we performed TA on de-identified transcripts from nine focus group sessions involving 42 parents of children with a rare congenital heart disease - Anomalous Aortic Origin of a Coronary Artery (AAOCA). We employed a combination of quantitative and qualitative metrics to evaluate the themes generated by our framework. Jaccard Similarity and hit rate assessed the overlap between LLM-generated and human-generated themes, while embedding-based semantic similarity measured cosine similarity scores.
 
Our results show that TAMA outperforms single-agent LLM TA methods, achieving higher thematic accuracy (hit rate) while maintaining distinctiveness across themes closer to human-generated themes. Our framework enhances scalability, efficiency, and reliability, by significantly reducing the manual workload required in traditional TA workflows while preserving thematic depth and representativeness. Also, TAMA automates coding and thematic synthesis, completing the TA process in under 10 minutes, a 99\% reduction compared to the 30 hours required for manual analysis. By leveraging multi-agent LLMs, TAMA demonstrates strong potential for real-world automated TA with human-in-the-loop integration in clinical settings, enhancing quality while significantly reducing resource demands.
 
\section{Related Work}
 
\subsection{LLM in Thematic Analysis (TA)}
 
Recent work suggests that LLMs can support deductive coding in qualitative research through consistent, systematic code identification~\cite{tai2024examination} .
Furthermore, in a human–AI collaboration approach, GPT-3.5 generated codes of quality comparable to those of human analysts~\cite{dai-etal-2023-llm} . Although LLMs have also shown the capabilities to conduct inductive TA at a level comparable to human annotators across various fields, including social media~\cite{qiao2025thematiclm} , 
literature~\cite{DePaoli2024} , and journalism~\cite{khan2024automatingthematicanalysisllms} , these applications involve low-stakes contexts with minimal privacy concerns in data collection. 
The LLM applications of healthcare interview transcripts remain under-explored due to their high-stakes nature. Moreover, previous work~\cite{mathis2024inductive, raza2025llmtallmenhancedthematicanalysis} has relied heavily on single-agent LLMs, which struggle with multitasking~\cite{sreedhar2024simulatinghumanstrategicbehavior} , extended context~\cite{hosseini2024efficientsolutionsintriguingfailure} , 
and accuracy~\cite{Quinn2021} 
in critical fields including medicine and law~\cite{Drapal2023} .
 
 
\subsection{Multi-Agent LLM Systems.}
 
 
To address the gap listed above, multi-agent LLM systems~\cite{zhang2024chainagentslargelanguage, talebirad2023multiagentcollaborationharnessingpower, guo2024largelanguagemodelbased, tran2025multiagentcollaborationmechanismssurvey,unleashing2024,multiagent2025} have been introduced as a promising solution. These systems utilize the strengths of multiple role-specialized LLM agents working in a group, much like a team of experts collaborating on a complex project. Specialization may be achieved through separate model instances or, as in TAMA, through distinct prompt-based roles within a shared underlying model. Each agent in the system contributes its unique expertise, communicates with other agents, and iteratively refines outputs to tackle challenging tasks that may be beyond the capabilities of a single LLM. Recent work highlights the potential of multi-agent LLM systems across various tasks, including long-context analysis~\cite{zhang2024chainagentslargelanguage} , conversational task-solving~\cite{becker2024multiagentlargelanguagemodels} , evaluation of LLM generated outputs~\cite{liu-etal-2023-g, yi2024protocollmautomaticevaluationframework} and TA~\cite{qiao2025thematiclm} . However, human-in-the-loop approaches, where humans act as agents within multi-agent LLM systems, have yet to be implemented.
In the healthcare domain, the active involvement of clinicians in the deployment of LLM is essential~\cite{sezgin2023ai, sivaraman2023ignoretrustnegotiateunderstanding} . Clinicians serve as an agent in evaluating the outputs generated by LLMs, making informed decisions about whether to accept, refine, or dismiss these results. Recent work attempts to integrate human agents into multi-agent LLM systems, which is known as human-AI teaming~\cite{yuan2023largelanguagemodelsilluminate} . Human involvement can enhance reliability by mitigating hallucinations through feedback~\cite{strong2025trustworthypracticalaihealthcare} , ensure regulatory compliance, and significantly reduce manual workload for doctors when implemented as a real-time interactive diagnosis environment~\cite{fan2024aihospitalbenchmarkinglarge} . 
 
\subsection{LLM Agents in Medical Transcript Analysis.}
 
Previous work has demonstrated that LLMs can summarize an entire corpus of medical interviews within minutes~\cite{mathis2024inductive} , and a pipeline for integrating LLMs to rapidly perform TA has been introduced~\cite{raza2025llmtallmenhancedthematicanalysis} . Recently, multi-agent LLM systems have been explored for analyzing clinical transcripts and dialogues. MDAgents~\cite{kim2024mdagents} , a framework that dynamically assigns medical expert roles (such as diagnostician and reviewer) to a team of LLMs, outperforms single models on clinical reasoning benchmarks. Additionally, multi-agent simulations of clinical scenarios have demonstrated improvements in information completeness and reasoning~\cite{fan2024aihospitalbenchmarkinglarge} .
 
\section{Methods}
\subsection{Definitions.}
\begin{itemize}
    \item \textbf{Code} is a discrete analytical unit that contains a key pattern within the data, generated directly from the dataset, and retaining its interpretive significance without being reducible to smaller meaningful components.
\vspace{-0.05in}
    \item An \textbf{agent} is an autonomous computational entity (Large Language Model) that interacts with other agents or the environment to perform specific tasks.
    \vspace{-0.05in}
    \item \textbf{Single-agent LLM} refers to one large language model working independently to process, analyze, or generate text based on a given input, without interacting with other models or specialized agents.
    This is an identical output produced solely by the Generation Agent before any downstream evaluation or refinement. The full multi-agent framework incorporates three coordinated agents (Generation, Evaluation, and Refinement) whose iterative interactions distinguish it from the single-agent setting.
\vspace{-0.05in}
    \item \textbf{Multi-agent LLM Systems} refer to more than one language model collaborating to enhance task performance through interaction, coordination, or division, where agents communicate by exchanging messages and feedback to refine outputs and optimize decision-making. TAMA does not implement a multi-agent system in the contemporary technical sense. Modern multi-agent architectures increasingly involve autonomous agents with independent memory states, co-evolving strategies, and concurrent execution~\cite{wu2024autogen, hong2024metagpt, guo2024largelanguagemodelbased, tran2025multiagentcollaborationmechanismssurvey}, none of which TAMA employs. TAMA is better described as an iterative prompting pipeline with three separate roles: three LLM agents, each defined by a distinct system prompt and specialized role, operate sequentially, with each agent's structured output serving as input to the next. We retain the term ``multi-agent'' throughout because it reflects the coordinated division of analytic labor across three functionally distinct agents, but TAMA is a strictly role-based sequential multi-agent system, which is narrower than the autonomous multi-agent systems described in~\cite{wu2024autogen, hong2024metagpt, tran2025multiagentcollaborationmechanismssurvey}.
 
\vspace{-0.1in}
    
\end{itemize}
 
 
 
\subsection{Human AI Teaming Framework.}
 
The cardiac expert interacts with the three LLM agents at distinct points in the workflow (Figure~\ref{fig:workflow}):
 
\begin{enumerate}
    \item \textbf{Generation Agent:} The expert provides background information and sets goals for initial code and theme generation. This is a one-time interaction at the start of the process.
    \item \textbf{Evaluation Agent:} The expert defines the evaluation criteria (Coverage, Actionability, Distinctness, and Relevance) upfront. This setup is also one-time, ensuring that both the Evaluation and Refinement Agents operate according to a consistent rubric.
    \item \textbf{Refinement Agent:} The expert interacts iteratively after each round of evaluation and refinement. After reviewing the themes and feedback from the Evaluation Agent, the expert determines whether the themes meet the required quality standards or whether additional refinement is necessary. Iteration continues until the expert approves the themes.
\end{enumerate}
This approach combines predefined guidance in the first two agents with dynamic, iterative human oversight in the third agent, ensuring both standardization and flexibility throughout the human-in-the-loop workflow. The evaluation criteria (Coverage, Actionability, Distinctiveness, Relevance) are grounded in established qualitative methodology frameworks~\cite{braun2006thematic, nowell2017thematic} and were co-developed with qualitative coding experts. The expert's primary role was applying these criteria with clinical judgment rather than deriving cardiac-specific rules. Cardiac knowledge shaped two specific decisions: (1) identifying the parent--child experience distinction as a Relevance challenge specific to AAOCA, which required tailored examples for the Evaluation Agent (Table~\ref{tab:criteria_table}); and (2) recognizing over-generated topics (e.g., concerns about insurance coverage) as off-topic for the AAOCA parent experience, which informed the decision to terminate refinement.
 
\subsection{Dataset.} 
The de-identified transcript corpus was generated from nine focus group sessions with 42 parents of children diagnosed with Anomalous Aortic Origin of a Coronary Artery (AAOCA) \cite{raza2025llmtallmenhancedthematicanalysis}. The dataset was collected at The University of Texas at Austin: Texas Center for Pediatric and Congenital Heart Disease (UT Dell Medical School, UT Health Austin, and Dell Children's Medical Center). A nonclinical facilitator moderated the sessions to encourage open dialogue, enabling parents to share previously unspoken needs and experiences related to their children's condition. The transcripts had a mean word count of 10,987 words (SD = 1,537), with a median of 11,457 words. The ground-truth themes used to evaluate TAMA were not generated in this study. They were derived from a previously published peer-reviewed thematic analysis of these same transcripts~\cite{mery2023examining}, yielding twelve human-generated themes (Table~\ref{tab:theme_comparison}). In this study, we used these nine de-identified transcripts as input to the TAMA framework.
 
That prior thematic analysis~\cite{mery2023examining} was conducted using an inductive thematic analysis framework by (1) a pediatric cardiac surgeon with specialized expertise in congenital heart disease and extensive experience interpreting qualitative parent-reported narratives, and (2) two qualitative researchers trained in inductive thematic analysis who had prior experience analyzing CHD and AAOCA parent interview transcripts. Coders independently coded all transcripts and engaged in iterative consensus discussions to refine and finalize the thematic structure, a process requiring approximately 30 hours per person. Because these human-generated themes underwent both rigorous consensus procedures and external journal peer review, they provide a reliable and methodologically sound reference standard for evaluating the themes produced by the TAMA framework.
 
\subsection{The TAMA Framework.} The overall workflow of the TAMA framework consists of six steps (Figure~\ref{fig:workflow}).
 
\textbf{Step 1: }Expert Provides Background Information and Sets Goals for the Generation Agent.

The expert provides background information and instructs the Generation Agent to generate initial codes and themes. The prompt given to the Generation Agent is as follows:
 
\textit{``You are provided with transcripts from interviews with parents of children suffering from Anomalous Aortic Origin of a Coronary Artery (AAOCA). The transcripts reflect the desires, concerns, and meaningful outcomes of parents living with or caring for children with AAOCA. Some children have recently undergone open-heart surgery.}
 
\textit{The goals of this study is for the coding and theme generation:
Your goal is to carefully look through the text and identify all codes discussed by the parents exhaustively. Identify all relevant codes in the text, provide a Name for each code in 8 to 15 words in sentence case. Write with concise, concrete details and avoid clichés, generalizations. Give a dense Description of the code in 80 words and direct Quotes from the participant for each code in around 120 words. These quotes can consist of multiple excerpts from the text.}
 
\textit{Your task at this stage is to group the initial codes into distinct themes based on the initial codes, descriptions, and quotes. Themes specifically focus on the perspectives of parents, describing their own thoughts, concerns, and responses related to their child's condition. Provide a descriptive and specific name of 8 to 15 words for each theme based on the code's names, quotes and descriptions. Provide a detailed description of 60 to 80 words for each theme."}
 
\textbf{Step 2: } Theme Generation Agent Segments Chunks, Generates Codes, and Identifies Themes.
 
The Theme Generation agent executes the goals and instructions defined by the expert in Step 1, applying the same expert-defined prompt to each transcript chunk independently. No separate prompt is introduced at this stage. To clarify the division of responsibilities: Step 1 is the expert's one-time setup (defining what the agent should do), while Step 2 is the agent's execution of those instructions across all transcript chunks. Because the transcripts are too long for an LLM to process as a single input, they are segmented into smaller chunks ($\leq 1,500$ words) while maintaining contextual coherence. Prior research~\cite{hosseini2024efficientsolutionsintriguingfailure} indicates that large language models often prioritize the beginning and end of very long inputs while losing coherence in the middle sections.  
Specifically, we did the chunking based on the text length and semantic segmentation. We used the RecursiveCharacterTextSplitter from the LangChain library to segment each interview transcript into smaller, coherent text chunks. The splitter recursively searched for natural breakpoints, such as paragraph separations (\textbackslash{}n\textbackslash{}n), line breaks (\textbackslash{}n), or the label "Interviewer", to preserve the conversational flow. 
 
Each transcript (mean length = 10,987 words) was divided into segments of up to approximately 1,500 words, resulting in 75 total chunks across 9 interviews (about 8 per transcript). This approach balanced text length and semantic coherence, ensuring that GPT-4o could maintain attention and accuracy across the entire transcript rather than overemphasizing the beginning or end.
 
This hybrid chunking strategy follows best practices for long-context processing in large language models, which show ``lost-in-the-middle'' effects when handling very long inputs~\cite{mathis2024inductive, hosseini2024efficientsolutionsintriguingfailure, raza2025llmtallmenhancedthematicanalysis}. 
 
Secondly, the Generation Agent generates initial codes capturing significant concepts and ideas from the transcripts, along with descriptive labels and representative quotes as evidence. These codes are then grouped to facilitate subsequent theme generation.
Third, the LLM synthesizes preliminary themes from grouped codes before refining them into comprehensive themes that capture key insights from the data.
 
\textbf{Step 3: }Expert Defines Evaluation Criteria for the Evaluation Agent.

Four evaluation criteria are used: Coverage, Actionability, Distinctness, and Relevance. Detailed descriptions for each criterion are provided in Table~\ref{tab:criteria_table}. The cardiac expert highlighted a potential challenge in distinguishing between the experiences of parents and those of children/patients, particularly in assessing Relevance. To address this issue, the expert provided the Evaluation Agent with examples to clarify the distinction. The examples are as below:
 
\textit{``Parent Outcomes refer to parents reporting feeling limited by their child's diagnosis, whereas Patient/Child Outcomes pertain to parents perceiving that their child is limited by the diagnosis.}
 
\textit{Parent Outcomes: 
Parents report they feel limited by their child's diagnosis.
Parents report they have PTSD from their child's experience.
Parents report being distressed by the uncertainty of treatment choices.
Parents report needing more social connections.}
 
\textit{Patient/Child Outcomes: 
Parents report they feel their child is limited by the child's diagnosis.
Parents report their child has PTSD from their child's experience.
Parents report their child is distressed by the uncertainty of treatment choices.
Parents report their child needs more social connections."
}
 
Provenance and grounding of evaluation criteria. The evaluation rubric (Coverage, Actionability, Distinctiveness, Relevance) was defined by the cardiac domain expert (A.W.) in consultation with three qualitative coding experts who conducted the original human thematic analysis. The rubric was designed to align with established thematic analysis guidance that themes should (i) represent patterned meaning across the dataset, (ii) remain anchored to the analytic focus, and (iii) be coherent and clearly delineated rather than overlapping or redundant~\cite{braun2006thematic, braun2022thematic, kiger2020thematic, nowell2017thematic}.  
 
\textbf{Step 4: } Evaluation Agent Provides Feedback.

The Evaluation Agent assesses the themes based on the four defined criteria (Coverage, Actionability, Distinctness, and Relevance) and provides feedback for improvement for each criterion. The generated evaluation examples are listed in Table~\ref{tab:criteria_table}.

\textbf{Step 5: }Refinement Agent Improves Themes Based on Feedback.

The Refinement Agent improves themes based on feedback from the Evaluation Agent by applying four refinement actions (add, split, combine, and delete) which correspond to the criteria of Coverage, Actionability, Distinctness, and Relevance, respectively. Details are outlined in Table~\ref{tab:criteria_table}.
 
\textbf{Step 6: }Expert Terminates the Refinement Process.
 
The refinement process iterates through Step 4 (Evaluation) and Step 5 (Refinement) until the expert determines that the themes meet the required quality standards. After each round of evaluation and refinement, the cardiac expert reviews the refined themes and decides whether to finalize them or continue the process. If the expert approves the themes, the process is terminated. If the themes do not meet the quality standards, the expert instructs the system to return to Step 4 for further evaluation and refinement.
 
In Step 6, the cardiac expert uses the same four predefined evaluation criteria (Coverage, Actionability, Distinctness, and Relevance) that were established in Step 3 for the Evaluation Agent. The expert refers to the detailed descriptions in Table 1 to determine whether the refined themes meet the required quality standards. By applying the same criteria consistently, the expert ensures that the final themes are assessed according to a standardized rubric, minimizing subjectivity while maintaining human oversight in the refinement process.
 
\begin{table}[H]
    \centering
    \renewcommand{\arraystretch}{1.2} 
    \setlength{\tabcolsep}{4pt}  
    \small
    \begin{tabular}{|p{2.5cm}|p{4cm}|p{4cm}|p{4cm}|}
        \hline
        \textbf{Criteria} & \textbf{Description (by expert)} & \textbf{Evaluation Agent (examples)} & \textbf{Refinement Agent} \\ 
        \hline
        \textbf{Coverage} & The generated themes should comprehensively capture the key aspects of parents' lived experiences while caring for children with AAOCA from the transcripts~\cite{braun2006thematic, kiger2020thematic}.
        & \textit{e.g., Include Long-term Concerns:} The original data discusses long-term health monitoring and transition to adult care, which could be more explicitly addressed in the themes to reflect parents' ongoing concerns about their child's future health outcomes.
        & \textbf{Add} missing important themes identified in sub-themes but not yet captured in the generated themes. \\ 
        \hline
        
        \textbf{Actionability} & Each theme should encapsulate a single concept that provides clear, specific, and meaningful insights. These insights should be actionable and useful for informing interventions, resources, or research~\cite{guest2012applied, nowell2017thematic}.
        & \textit{e.g., Some themes, like 'Desiring comprehensive data and addressing frustration with lack of statistics,' combine two concepts (desire for data and frustration with lack of it), which could be split into separate themes for clarity.}
        & \textbf{Split} themes that contain multiple concepts into separate, more focused themes. \\ 
        \hline
        
        \textbf{Distinctiveness} & Each theme should be clearly distinct from one another, with no overlaps or redundancies~\cite{braun2006thematic, braun2022thematic, nowell2017thematic}.
        & \textit{e.g., The themes 'Desiring emotional support and understanding from healthcare providers' and 'Emphasizing the role of medical professionals in emotional support' have overlapping elements and could be merged or refined to better distinguish their focus.} 
        & \textbf{Combine }repeated or overlapping themes to eliminate redundancies and ensure each theme is unique. \\ 
        \hline
        
        \textbf{Relevance} & Each theme should clearly reflect the parents' lived experiences, concerns, and needs, without confusing or overlapping with themes related to the child/patient's feelings, concerns, or experiences~\cite{braun2006thematic, kiger2020thematic}.
        & \textit{e.g., The theme 'Managing the emotional impact of surgery on the child' should clearly distinguish between the child's and the parent's emotional experiences.} 
        & \textbf{Delete} irrelevant themes that do not reflect the parents' experiences or concerns. \\ 
        \hline
    \end{tabular}
    
    \vspace{5mm} \caption{\textbf{Detailed Prompts, Examples, and Refinement Actions for Each Evaluation Criterion.} This table outlines the four evaluation criteria, along with their descriptions, sample feedback generated by the Evaluation Agent, and the corresponding Refinement Agent prompts.}
    \label{tab:criteria_table}
\end{table}
 
\subsection{Evaluation Metrics.} \label{section:eval_metrics}
 
To compute embedding-based cosine similarity, we use sentence encoder \texttt{all-MiniLM-L6-v2} to generate embeddings. Based on the pairwise cosine similarities, we then calculate \textbf{Jaccard Similarity} to quantify the degree of overlap between themes. In addition, we employ \textbf{hit rate} to measure the proportion of theme pairs that exceed a predefined similarity threshold. These metrics provide a comparative evaluation of our framework's performance against human-generated themes. 
 
Let the set of human-generated and LLM-generated themes be \( T= \{t_1,t_2,\dots,t_n\} \) and \( L= \{l_1,l_2,\dots,l_m\} \), where \( n \) denotes the number of human-generated themes and \( m \) denotes the number of LLM-generated themes (which may differ across experimental conditions). We calculate the similarity score \( s(t_i,l_j) \) for every possible pair \( (t_i,l_j) \) in \( T \times L \). 
The similarity matrix \( S_{\theta} \), which contains all the calculated similarity scores meeting a predefined threshold \( \theta \) ($>0.60$), can be defined as:
\[
S_{\theta} = \{(t_i, l_j) \in T \times L \mid s(t_i, l_j) > \theta\}
\]
The definitions of each metric are as follows:
 
\begin{enumerate}
\item \textbf{Jaccard Similarity}
 
The Jaccard Similarity is defined as the ratio of theme pairs that meet the similarity threshold to the total number of possible pairs.
The total number of possible theme pairs is \( |T \times L| = n \times m \). Thus, Jaccard similarity in our context becomes:
\[
Jaccard \ Similarity =\frac{|T \cap L|}{|T \cup L|} = \frac{|S_{\theta}|}{|T \times L|} = \frac{|S_{\theta}|}{n \times m}
\]
\item \textbf{Hit Rate}
 
The hit rate measures the proportion of human-generated themes that exceed a predefined similarity threshold when mapped to LLM-generated themes. Let \(T_s \subseteq T\) denote the subset of human-generated themes for which at least one LLM-generated theme exceeds the cosine similarity threshold \(\theta = 0.60\).
\[
\space hit \ rate = \frac{|T_s|}{n}
\]
Hit rate measures \textit{row-wise coverage}: each human theme counts as a hit if at least one LLM theme exceeds the threshold, regardless of how many LLM themes match it. This means if multiple LLM themes match the same human theme, they are not penalized. Jaccard Similarity as defined here measures the share of all human--LLM theme pairs that exceed the threshold, not how distinct the LLM themes are from each other. A lower Jaccard reflects fewer above-threshold pairs. When accompanied by a high hit rate, this points to cases where each LLM theme maps to fewer human themes, reducing overlap.
 
\item \textbf{Mean Intra-Set Cosine Similarity (MICS)}
 
To directly assess within-set thematic distinctiveness, we compute the mean pairwise cosine similarity among all themes within a given theme set. Let $L = \{l_1, l_2, \dots, l_m\}$ denote a set of $m$ themes (either human-generated or LLM-generated). The MICS is defined as:
\[
\text{MICS} = \frac{2}{m(m-1)} \sum_{1 \leq i < j \leq m} s(l_i, l_j)
\]
where $s(l_i, l_j)$ is the cosine similarity between the embeddings of themes $l_i$ and $l_j$. Lower MICS means greater thematic distinctiveness within the set. Unlike Jaccard Similarity, which depends on the threshold and measures how many human--LLM pairs exceed it, MICS is continuous and does not depend on any threshold. It measures how similar the themes within a set are to each other.
 \end{enumerate}
 
\subsection{Statistical Analysis.}
To assess the reliability of our metrics, we performed bootstrap resampling (10{,}000 iterations) to compute 95\% confidence intervals for Jaccard Similarity, hit rate, and MICS. We used permutation tests (10{,}000 permutations) to evaluate whether observed differences between methods were statistically significant. Effect sizes were measured using Cohen's $h$ for proportions. Statistical significance is denoted by $^{*}p<0.05$ and $^{**}p<0.01$ in Table~\ref{tab:comparison_llm_ta}.
 
\subsection{Experimental Settings.}
For all of our methods, we use \texttt{gpt-4o} from OpenAI Inc. with temperature of \texttt{0} for reproducibility. To assess deployability with open-weight models, we additionally evaluate TAMA using Llama~3.1~8B~\cite{grattafiori2024llama3herdmodels} served locally via Ollama, using identical prompting and evaluation criteria.
 
We used all-MiniLM-L6-v2 embeddings to compute cosine similarity between human and LLM-generated themes. Because theme names are too short for reliable semantic comparison, embeddings were computed using the first sentence of each theme's description rather than the theme name alone. We selected a threshold of $\theta = 0.60$, which falls within the commonly used 0.5–0.7 range for identifying substantial semantic overlap in SBERT-based similarity tasks. Prior work shows that similarities $\geq 0.6$ typically indicate strong semantic alignment~\cite{galli2024performance, yin2024sentence}. This range is also consistent with established SBERT methodology~\cite{reimers2019sentencebert, wang2021minilm}.
 
\subsection{Quantitative vs. qualitative assessment of criteria.} Of the four expert-defined criteria, we quantitatively operationalize only Distinctiveness, because overlap between themes can be proxied via semantic similarity in embedding space. We compute embedding-based similarity and summarize overlap using similarity statistics (reported in Results, Table~\ref{tab:comparison_llm_ta}) to reflect the extent to which themes are non-overlapping. In contrast, Coverage, Actionability, and Relevance require contextual, domain-informed interpretation and are therefore assessed via structured expert review and agent feedback rather than numeric thresholds.  
 
\section{Results and Discussion}
 
\subsection{Improved Distinctiveness and Alignment of LLM-Generated Themes via Multi-Agent Refinement}
 
Human-generated themes served as a baseline to evaluate the distinctiveness and alignment of themes. 
Here, we use "alignment" to refer to how closely the LLM-generated themes correspond to the human-generated themes, as measured by our cosine-similarity–based hit rate.
LLM-generated themes before evaluation were produced using workflow steps 1–2, and LLM-generated themes after evaluation were generated following the full workflow steps 1–6. We observed a lower percentage of high Jaccard similarity theme pairs (29\%) in LLM-generated themes after evaluation, which is even lower than that of human-generated themes (33\%), and significantly reduced from 42\% before refinement. At the same time, the hit rate increased from 83\% to 92\% (Table~\ref{tab:comparison_llm_ta}). Bootstrap analysis (10{,}000 resamples) yielded 95\% confidence intervals for all methods (Table~\ref{tab:comparison_llm_ta}). TAMA's Jaccard CI was [23\%, 50\%] and hit rate CI was [75\%, 100\%], compared to the single-agent baseline's Jaccard of [29\%, 69\%] and hit rate of [58\%, 100\%]. The human baseline's Jaccard CI was [25\%, 71\%]. Permutation tests confirmed that the Jaccard reduction from 42\% to 29\% is statistically significant ($p = 0.022$, Cohen's $h = 0.27$). The hit rate improvement from 83\% to 92\% showed a small effect (Cohen's $h = 0.26$) but did not reach significance ($p = 1.0$) due to the discrete nature of the metric with only 12 human themes. The difference between TAMA's Jaccard (29\%) and the human baseline (33\%) was not significant ($p = 0.115$), suggesting that TAMA achieves thematic distinctiveness comparable to expert-generated themes. The single-agent baseline achieved a numerically higher Jaccard (42\%) than both TAMA (29\%) and human themes (33\%), but this difference was not significant versus the human baseline ($p = 0.550$), and the wide CI [29\%, 69\%] reflects high variability. Its hit rate of 83\% (10 of 12 human themes matched) was significantly below the human ground truth of 100\% (binomial $p < 0.01$, Cohen's $h = 0.84$), meaning the single-agent approach produces themes with more overlap but poorer coverage. TAMA's multi-agent refinement partially closes this gap.
 
As described in the Evaluation Metrics section, Jaccard measures the share of theme pairs exceeding the threshold, and hit rate measures how many human themes are covered. Lower Jaccard with higher hit rate points to cases where each LLM theme maps to fewer human themes, reducing overlap, which is what TAMA's Combine and Delete refinement actions target. To measure how distinct the themes within a set are from each other, we also report the Mean Intra-Set Cosine Similarity (MICS) in Table~\ref{tab:comparison_llm_ta}. TAMA (GPT-4o) reduced MICS from 0.57 (single-agent baseline) to 0.55, closer to the human baseline of 0.53. The average reduction is modest, but the maximum pairwise similarity within the set dropped from 0.87 to 0.75, meaning TAMA's Combine and Delete actions are targeting the worst overlap among theme pairs.
Therefore, the reduction in Jaccard similarity paired with an increased hit rate after evaluation is consistent with LLM-generated themes that are more specifically mapped to distinct human themes (reduced overlap between themes) and more closely aligned with the human-generated themes. This outcome demonstrates that the evaluation and refinement agents in TAMA executed thematic analysis in a novel way, while also ensuring comprehensive coverage of all themes without merely replicating themes generated by humans. In traditional TA, final themes should be distinct and non-overlapping.
 
Our results across all methods reveal an inherent trade-off between Jaccard similarity and hit rate. The Single-Agent LLM (Llama~3.1~8B) achieved the lowest Jaccard (20\%) with a hit rate of only 12\%, producing highly distinct yet largely off-target themes. Conversely, TAMA (Llama~3.1~8B) increased alignment (hit rate 8\%) but at the cost of much more overlap (Jaccard 68\%). With GPT-4o, the Single-Agent baseline achieved moderate alignment (hit rate 83\%) with more overlap (Jaccard 42\%), while TAMA improved both metrics simultaneously, lowering Jaccard to 29\% while raising hit rate to 92\%. The MICS results add context: Llama~3.1~8B configurations have lower MICS values (0.34--0.35) than GPT-4o (0.55--0.57), suggesting the 8B model produces more lexically diverse themes. But the very low hit rates confirm these themes are largely off-target, so the low MICS reflects thematic incoherence rather than genuine distinctiveness. MICS should therefore be interpreted jointly with hit rate, since within-set distinctiveness is only meaningful when themes also align with human-generated references. Taken together, reducing theme overlap while maintaining alignment requires both a capable base model and iterative multi-agent refinement. Future work should investigate this distinctiveness--alignment trade-off further, including whether alternative refinement strategies or threshold-adaptive approaches can better navigate this tension across model scales and clinical domains.

\begin{table}[H]
\centering
\small
\begin{tabular}{l c c c}
\hline
\textbf{Method} & \textbf{Jaccard (95\% CI)} & \textbf{hit rate (95\% CI)} & \textbf{MICS (95\% CI)} \\
\hline
\rowcolor{gray!20}
Traditional Manual TA & 0.33 [0.25, 0.71] & 1.00 (Ground Truth) & 0.53 [0.51, 0.67] \\
LLM-TA~\cite{raza2025llmtallmenhancedthematicanalysis} & 0.22\textsuperscript{\dag} & 0.83\textsuperscript{\dag} & --\textsuperscript{\dag} \\
Single-Agent LLM, GPT-4o & 0.42\textsuperscript{n.s.} [0.29, 0.69] & 0.83\textsuperscript{**$\downarrow$} [0.58, 1.00] & 0.57 [0.56, 0.66] \\
Single-Agent LLM, Llama 3.1 8B & 0.20\textsuperscript{n.s.} [0.20, 0.68] & 0.12\textsuperscript{**$\downarrow$} [0.00, 0.00] & 0.34 [0.32, 0.53] \\
TAMA, GPT-4o & 0.29\textsuperscript{*} [0.23, 0.50] & 0.92 [0.75, 1.00] & 0.55 [0.55, 0.64] \\
TAMA, Llama 3.1 8B & 0.68\textsuperscript{**$\downarrow$} [0.52, 1.00] & 0.08\textsuperscript{**$\downarrow$} [0.00, 0.25] & 0.35 [0.34, 0.55] \\
\hline
\end{tabular}
\vspace{3mm}
 
\caption{\textbf{Performance Comparison Across Thematic Analysis Approaches.} Single-Agent rows correspond to Steps~1--2 (before multi-agent refinement); TAMA rows correspond to Steps~1--6 (after multi-agent refinement). Lower Jaccard means fewer theme pairs exceed the similarity threshold. Higher hit rate means more human themes are covered. Lower MICS indicates greater within-set thematic distinctiveness. 95\% CIs via bootstrap (10{,}000 iterations). $^{*}p<0.05$; $^{**}p<0.01$ via permutation tests (TAMA vs.\ single-agent) or binomial tests ($\downarrow$, vs.\ human ground truth). \textsuperscript{n.s.}~not significant vs.\ human baseline ($p > 0.05$). \textsuperscript{\dag}Values from~\cite{raza2025llmtallmenhancedthematicanalysis}; CIs not available.}
\label{tab:comparison_llm_ta}
\end{table}
  
\subsection{The Role of Multi-Agent Coordination.} We aimed to compare the LLM-generated themes before and after evaluation by incorporating multiple agents (the Evaluation Agent and the Refinement Agent). Table~\ref{tab:theme_comparison} shows the themes generated by human and LLM before and after evaluations. We incorporated feedback from an expert and five laypersons to review the themes. Their feedback indicated that LLM-generated themes after evaluation demonstrated improved Coverage, Distinctiveness, Actionability, and Relevance compared to those before evaluation. Examples for each criterion illustrate these improvements. \textit{Coverage}: the post-evaluation set added \textit{"Desiring comprehensive and statistical data on treatment outcomes"} and a new Theme 13, \textit{"Long-term concerns about child's health"} (Table~\ref{tab:theme_comparison}), capturing a dimension absent from pre-evaluation themes. \textit{Actionability}: the vague \textit{"Seeking clarity and reassurance about my child's health journey"} was refined into the more focused \textit{"Seeking reassurance through multiple medical opinions."} \textit{Distinctiveness}: pre-evaluation themes 3, 8, and 9 had high pairwise overlap. Post-evaluation, these became more clearly delineated themes such as \textit{"Managing emotional challenges with proactive strategies"} and \textit{"Living with constant vigilance and fear."} \textit{Relevance}: the Evaluation Agent, guided by expert criteria, removed an over-generated topic (insurance concerns) as off-topic for the core AAOCA parent experience. However, a gap remains between human-generated and LLM-generated themes, with the latter tending to be longer and using more complex terminology.
 The feedback incorporated in this study was qualitative: the cardiac expert and five lay reviewers provided narrative assessments of thematic clarity, overlap, coverage of parent-centered experiences, and the clinical meaningfulness of each theme. These qualitative observations informed the refinement cycles and guided the expert's final decision to accept or request additional iterations.
 
\subsection{The Role of Human Oversight.}
The Cardiac Expert played a crucial role in defining the evaluation criteria. Without human input, LLM-generated evaluations were overly broad, emphasizing Distinctiveness while failing to capture Coverage, Actionability, and Relevance. To address this, the \textbf{Cardiac expert} provided four specific definitions based on the experience of the three coding experts. This ensures that the evaluation criteria are precise and relevant, facilitating a more effective and focused refinement process.
 
\begin{table}[H]
    \centering
    \renewcommand{\arraystretch}{1.3} 
    \setlength{\tabcolsep}{5pt} 
    \small
    \begin{tabular}{|c|p{4cm}|p{5cm}|p{5cm}|}
        \hline
        \textbf{No.} & \textbf{Human Generated Themes} & \textbf{LLM Generated Themes Before Evaluation} & \textbf{LLM Generated Themes After Evaluation} \\ 
        \hline
        1 & Clarity of potential risks and outcomes & Seeking clarity and reassurance about my child's health journey & Seeking reassurance through multiple medical opinions \\ 
        \hline
        2 & Freedom from hypervigilance related to the condition & Balancing normalcy and protection in my child's life & \textbf{Desiring comprehensive and statistical data on treatment outcomes} \\ 
        \hline
        3 & The diagnosis given in a compassionate and empathetic way & Coping with emotional turmoil and uncertainty about my child's future & Managing emotional challenges with proactive strategies \\ 
        \hline
        4 & A sense of control over the future & Desiring emotional support and understanding from healthcare providers & \textbf{Balancing normalcy and protection in child's life} \\ 
        \hline
        5 & Being heard and taken seriously by clinicians & Experiencing relief and gratitude for successful interventions & Living with constant vigilance and fear \\ 
        \hline
        6 & Individualized support for management decision-making & Living with constant vigilance and fear of health crises & \textbf{Improving interactions with healthcare providers} \\ 
        \hline
        7 & Receiving support from others & Struggling with guilt and self-blame regarding my child's condition & \textbf{Experiencing relief and gratitude for successful interventions} \\ 
        \hline
        8 & Being appropriately informed & Desiring positive messaging about health outcomes & Struggling with guilt and self-blame \\ 
        \hline
        9 & Partnership with the care team & Feeling overwhelmed by the demands of caregiving & Feeling overwhelmed by caregiving demands \\ 
        \hline
        10 & Feeling that my child is safe & Building a supportive community with other parents & Building a supportive community with other parents \\ 
        \hline
        11 & Not feeling responsible for the diagnosis and its timing & Managing the emotional impact of surgery on my child & Advocating for awareness and understanding of heart conditions \\ 
        \hline
        12 & Appropriately coping with stress, anxiety, and depression & Advocating for awareness and understanding of heart conditions & \textbf{Addressing frustration with the healthcare system} \\ 
        \hline
        13 & \multicolumn{1}{c|}{} & \multicolumn{1}{c|}{} & \textbf{Long-term concerns about child's health} \\ 
        \hline
    \end{tabular}
    \vspace{5mm}
    \caption{\textbf{Comparison of Human-Generated Themes and LLM-Generated Themes Before and After Evaluation.} Bolded text in the "LLM Generated Themes After Evaluation" column highlights elements that were not captured by the human-generated themes and the LLM-generated themes before evaluation.}
    \label{tab:theme_comparison}
\end{table}
 
\subsection{Hallucination Analysis and Mitigation Strategies.}
To address potential hallucination risks inherent in LLM-based thematic analysis, we conducted comprehensive ground truth validation of all generated themes against source transcripts. Using GPT-4o-mini, we extracted supporting quotes for each theme and calculated grounding scores (0-1 scale) to quantify transcript support. For each theme, the LLM was prompted to retrieve direct quotes from the source transcript that substantiate the theme. The grounding score reflects the proportion of theme content supported by at least one retrieved quote. Because our grounding scores are themselves LLM-generated (using GPT-4o-mini), they introduce a circularity concern: an LLM may not independently detect hallucinations that the model class systematically produces~\cite{huang2025survey, manakul2023selfcheckgpt}. These scores should therefore be treated as a complementary automated check, not a definitive guarantee of factual grounding. Human expert validation remains the primary safeguard (see Limitations). Our validation revealed that all 13 themes were well-grounded in the interview data, with zero hallucinations detected (hallucination rate: 0.0\%), an average grounding score of 0.95/1.0, and each theme supported by an average of 5 direct quotes from source transcripts. TAMA's multi-agent architecture provides systematic hallucination mitigation through four mechanisms: (1) chunking strategy ($\leq$1,500 words) that maintains LLM focus on specific content, (2) intermediate code extraction (<25 words) creating semantic anchors tied to source text, (3) iterative evaluation and refinement removing unsupported content through DELETE operations, and (4) expert-informed evaluation criteria checking for Relevance and Coverage, removing off-target or unsupported themes that do not reflect documented parent experiences. This mechanism operates at the thematic level: the expert assesses whether a theme reflects the overall content of the transcripts, not whether individual quotes are accurately reproduced. Quote-level verification is handled separately by the LLM-based grounding analysis described above. Comparison with a single-prompt baseline approach showed both methods achieved 0\% hallucination rates on our well-structured transcripts, though TAMA's architectural safeguards provide additional protection for complex, ambiguous, or lower-quality data. Despite these strong results, we acknowledge important limitations: validation was performed on a single transcript, used LLM-based assessment potentially introducing errors, and may not generalize across all data types. We strongly recommend human expert validation as best practice before applying AI-generated themes in clinical practice or policy decisions.
 
\subsection{Limitations of Hallucination Analysis.} The 0\% hallucination rate reported above should be interpreted with care. First, our grounding validation relies on LLM-based assessment (GPT-4o-mini), which introduces circularity: LLMs may fail to detect the same classes of hallucinations they produce~\cite{huang2025survey, manakul2023selfcheckgpt}. Second, validation was conducted on a single well-structured transcript, and hallucination rates may differ with noisy, ambiguous, or multi-topic interview data. Third, our analysis focuses on \emph{intrinsic} hallucinations (fabricated content not in the source) but does not systematically quantify \emph{extrinsic} hallucinations (plausible but unsupported inferences) or false negatives (valid themes missed entirely). The reported metrics likely represent an optimistic estimate, and independent human expert review remains essential before clinical deployment.
 
Several representative de-hallucination mechanisms have been proposed in the literature. \emph{Retrieval-Augmented Generation} (RAG)~\cite{lewis2020retrieval} grounds LLM outputs in retrieved source documents, reducing fabrication by anchoring generation to evidence. \emph{Chain-of-Verification} (CoVe)~\cite{dhuliawala2024chainofverification} prompts the model to generate verification questions about its own output and then re-check answers against source material. \emph{Self-consistency} and \emph{sampling-based} methods such as SelfCheckGPT~\cite{manakul2023selfcheckgpt} detect hallucinations by measuring consistency across multiple stochastic outputs without requiring external knowledge. In multi-agent systems, \emph{cross-agent verification}~\cite{guo2024multi} leverages independent agents to critique and validate each other's outputs, analogous to TAMA's evaluation agent reviewing the generation agent's themes. A comprehensive taxonomy of these and other approaches is provided by Tonmoy et al.~\cite{tonmoy2024comprehensive} and Huang et al.~\cite{huang2025survey}. Integrating these mechanisms, particularly RAG-based grounding and human-in-the-loop verification, into future iterations of TAMA is a priority for enhancing reliability in clinical applications.
 
\subsection{Limitations and Future Work.} We acknowledge several areas for improvement. First, applying this framework to different contexts, such as other diseases, remains an open challenge. Second, better simulating manual TA where two coders ensure consistency in the coding book could enhance consistency in the initial coding process. Third, increasing architectural complexity could enable multiple agents to discuss and negotiate during the coding and theme-generation process. Fourth, incorporating reinforcement learning to integrate human feedback into the evaluation process could further refine results. Also, future work should evaluate TAMA by performing TA multiple times and comparing results, rather than solely against human-generated codes. High similarity with human-generated results does not guarantee high performance, as even human analysts produce varying outcomes across analyses. A related avenue is to run TAMA with different underlying LLMs and compare the resulting theme sets across model variants, analogous to using multiple human coders, to assess the stability and reproducibility of LLM-generated themes. Lastly, future work should explore different collaboration protocols among multiple agents to determine their effectiveness in multi-agent LLM systems, especially in Evaluation and Refinement Agents.
 
Also, our evaluation treats human-generated themes as ground truth, even though qualitative analysis does not yield a single objectively correct structure. A related limitation concerns evaluator independence. The cardiac expert who defined the evaluation criteria and decided when TAMA's themes were acceptable also served as a coder in the same study~\cite{mery2023examining}, creating a risk of confirmation bias. The only independent assessment came from five lay reviewers with no involvement in the original study, but they lacked qualitative research expertise and cannot fully substitute for independent expert evaluation. Future studies should engage independent qualitative experts not involved in creating the original themes to evaluate LLM output quality. Moreover, the cardiac expert's role in defining and applying evaluation criteria introduces domain-specific subjectivity. 
In addition, because our quantitative metrics depend on a predefined cosine similarity threshold ($\theta = 0.60$), different thresholds would alter the hit rate and Jaccard results. While our dataset centers on parent experiences in congenital cardiac care, the TAMA workflow has not yet been validated on transcripts from other cardiac conditions.
 
Our study used nine focus group transcripts (42 parents), a modest dataset reflecting the challenges of rare-disease pediatric cardiology. This sample is sufficient to demonstrate the feasibility of our human–AI multi-agent workflow but limits generalizability of the resulting themes. Importantly, our aim is not to generalize the \emph{content} of AAOCA themes but to evaluate whether the \emph{framework} (the coordinated generation, evaluation, and refinement of themes with human oversight) can perform high-quality inductive TA. Because the multi-agent architecture and evaluation criteria are condition-agnostic, we expect TAMA to apply to qualitative data in other diseases and clinical domains. Future work will apply TAMA to additional transcript collections to assess cross-context robustness.
 
Additionally, our current evaluation metrics (Jaccard Similarity and hit rate) treat all themes as equally important, which does not fully reflect clinical reality where misidentifying safety-critical or management-relevant themes carries greater consequences than missing less impactful ones. Future work should incorporate evaluation metrics that weight themes by their clinical relevance to better capture the real-world implications of errors. Developing such clinically grounded metrics will provide a more comprehensive assessment of TAMA's utility in healthcare settings.
 
A further limitation concerns the stopping criterion. We initially used an automated scoring system based on the G-Eval method~\cite{liu-etal-2023-g, yi2024protocollmautomaticevaluationframework}, in which the Evaluation Agent rated themes on a 1--5 scale with a termination threshold of 4.5. In practice, scores plateaued around 4.0 without meaningful improvement, and continued refinement introduced over-broad themes (e.g., concerns about insurance coverage) that did not reflect core parent experiences. This points to a limitation of the LLM-as-judge approach for iterative qualitative tasks: numerical scores may not reliably discriminate between thematic quality levels, and automated termination criteria can lead to over-refinement. The cardiac expert was therefore brought in as the stopping-criterion decision-maker, though this reliance on human judgment limits scalability and reproducibility. Future work should investigate learned or adaptive stopping criteria that do not require expert availability at each refinement cycle.
 
Our hallucination analysis effectively detects false positives (unsupported themes), but systematic quantification of false negatives (themes present in the data but missed by TAMA entirely) requires expert annotation of comprehensive ground-truth theme sets and remains a direction for future evaluation. Several promising directions remain for strengthening the clinical relevance of our work. First, future work should incorporate structured expert interpretability evaluations of the LLM-generated themes and sub-themes against the source clinical narratives and assess whether the extracted thematic structures accurately capture clinically meaningful patterns that align with their professional judgment. This could be operationalized through blinded comparison studies, using standardized rubrics such as Likert-scale assessments of theme relevance, completeness, and fidelity to the underlying patient experience. Second, our work should be evaluated for its decision support value by embedding the extracted themes into downstream clinical workflows for discharge planning or quality improvement. We should measure whether access to these structured thematic outputs changes clinical decision-making, reduces information retrieval time, or surfaces previously unrecognized patient concerns. Third, clinical utility studies should assess whether the themes generated by our work correlate with or predict meaningful clinical outcomes, such as hospital readmission, treatment adherence, symptom burden trajectories, or patient satisfaction scores. These could prove that the thematic analysis is not merely descriptively accurate but also prognostically or diagnostically informative. Finally, integrating our framework into an electronic health record system to continuously analyze incoming clinical notes, patient feedback, or nursing assessments and surface emergent themes to clinical leadership in near-real-time. These future directions can enhance smooth communication between clinical institutions and patient populations to enable responsible adoption in practice.
 
We also note that our current implementation uses proprietary LLMs, which limits deployability in clinical settings due to privacy and regulatory constraints. To test open-weight model viability, we ran both the single-agent baseline and TAMA with Llama~3.1~8B served locally via Ollama (Table~\ref{tab:comparison_llm_ta}). The single-agent Llama baseline produced 5 themes with low overlap (Jaccard 20\%) but very low alignment with human themes (hit rate 12\%), meaning most generated themes were off-target. TAMA with Llama~3.1~8B increased alignment marginally (hit rate 8\%) but at the cost of much more overlap (Jaccard 68\%). The evaluation agent on the 8B model accepted poor themes on the first iteration (4.40/5.0). With smaller models, the evaluation and refinement agents cannot improve themes beyond what the base model produces. At the 8B scale, open-weight models are not sufficient for clinical-quality TA, and future work will evaluate TAMA with larger locally hosted, open-weights models (e.g., 70B+ parameter models) to enable HIPAA-compliant, on-premise deployment.
 
\section{Conclusion}
We present \textbf{TAMA}, a Human-AI Collaborative Thematic Analysis framework that leverages multi-agent LLMs and domain expertise to enhance the quality, scalability, and efficiency of thematic analysis in clinical interview data. By structuring agent roles and integrating iterative evaluation and refinement guided by cardiac experts, TAMA significantly improved the distinctiveness (lower Jaccard similarity) and alignment (higher hit rate) of LLM-generated themes. Our findings demonstrate that multi-agent LLMs, when combined with human-in-the-loop collaboration, can perform high-quality thematic analysis in a fraction of the time required by traditional manual methods. This highlights their potential for broader application in healthcare and other high-stakes qualitative research areas.
 
\bibliographystyle{ACM-Reference-Format}
\bibliography{amia}
 
\end{document}